\documentclass[12pt, preprint]{aastex}
\usepackage{natbib}
\begin{document}
\title{Can Ultrahigh Energy Cosmic Rays Come from Gamma-Ray Bursts? \\
II: Cosmic Rays Below the Ankle and Galactic GRB}
\author{David Eichler\altaffilmark{1},
 Martin Pohl\altaffilmark{2,3}}
\altaffiltext{1}{Physics
Department, Ben-Gurion University, Be'er-Sheva 84105, Israel;}
\altaffiltext{2}{Institut f\"ur Physik und Astronomie, Universit\"at
Potsdam, 14476 Potsdam-Golm, Germany}
\altaffiltext{3}{DESY, 15738 Zeuthen, Germany}

\begin{abstract}
The maximum cosmic ray energy achievable by acceleration by a relativistic blast wave is derived. It is shown that forward shocks from long GRB in the interstellar medium
are powerful enough to produce the Galactic cosmic-ray component
{ just below} the ankle at
$4\times 10^{18}$eV, as per an earlier suggestion \citep{le93}. { It is further  argued that, were extragalactic long GRBs  responsible for the component {\it above} the ankle as well,  the   occasional Galactic GRB within the solar circle  would contribute more than the observational limits on the outward flux from the solar circle, unless an avoidance scenario, such as intermittency and/or beaming, allows the present-day, local flux   to be less than $10^{-3}$ of the  average.  Difficulties with these avoidance scenarios are  noted.}
\end{abstract}


\section{Introduction}

The ultrahigh-energy (UHE) range of the cosmic-ray (CR) spectrum is generally broken into
three parts: 1) the steeper knee-to-ankle segment ($\sim 10^{15.5}$
to $10^{18.6} $eV), 2) the flatter "trans-ankle" CR  below the GZK cutoff  ($10^{19.6}$ eV), 3) and trans-GZK CR. Trans-ankle CR are probably  extragalactic, showing little anisotropy at $E\le10^{19.6}$ eV, and some
anisotropy at $E\ge10^{19.6}$ eV towards the local supercluster.

In this paper, we consider the hypothesis  of sub-ankle UHECR
origin from long GRB \citep{le93,wda04,2010PhRvL.105i1101C}. We show that (a) Galactic $\gamma$-ray bursts (GRB) are sufficiently good accelerators and sufficiently powerful to account for sub-ankle UHECR, but that (b) the UHECR near-isotropy limits the current  Galactic  UHECR output per unit star-forming mass to a value far less than what is energetically required to account for trans-ankle extragalactic UHECR by extragalactic GRB. Conclusion (b) can be generalized to any hypothetical UHECR source whose rate density, like that of long  GRB, is in  proportion  to star formation. This challenges any model in which such sources   account for {\it all} UHECR.

Many past authors have  proposed that  GRB make the CR above $10^{19}$ eV, but for this energy
range there remains the alternative hypothesis that they come from active galaxies
\citep{2010APh....34..314T}.
Doubts remain that GRB could supply the highest-energy cosmic rays. The problems
include disparity in the total energetics of each \citep[Part I, and
references therein]{egp10}, adiabatic losses, which lower the maximum energy
should the acceleration be in a compact region, and the isotropy problem.
The isotropy problem, discussed here and in an accompanying paper (Part III), is basically that stars in the Milky Way
are distributed anisotropically relative to the Earth. If they -
or sources similarly distributed -  were responsible for UHECR,
 the UHECR should also show anisotropy, the Galactic magnetic field notwithstanding.
Part III studies in detail the propagation of CR from Galactic GRB and
specifically compare with data the expected anisotropy, composition, and intermittency
behavior.

\section{Particle Acceleration in Outflows: A General Discussion}

To  be efficiently shock accelerated, a CR that has crossed
the shock toward the upstream must be overtaken again by the shock of the order of $  n \equiv \beta/\beta_S$ times, where $\beta$, the velocity of the particle in units of c, exceeds $\beta_S$, the velocity of the shock. If the shock
moves at Lorentz factor $\Gamma_S$, then a CR that is overtaken by a spherical blast wave of age T at radius $R_S\sim \beta_S c T$
must have been deflected (by gyration or scattering)  through an angle  $\Delta \theta\gtrsim 1/\Gamma_S$, while residing upstream, within
a time $\Delta t \sim R_S/\beta c n =  \beta_S  R_S/ \beta^2 c$.\footnote{Here it is assumed that, whether the propagation is stochastic or scatter-free, each reorientation or reversal of direction  must happen within a CR path length of order  $ \beta_S R_S/\beta$,   as the cumulative time a CR spends within a gyroradius $r_g$ of the shock is only $\sim R_S/\beta c$.  This assumption may be controversial for scatter-free propagation in a perpendicular magnetic field, because there is no rigorous proof to our knowledge of impossibility of more prolonged trapping, but neither are we aware of any counterexample in view of systematic drift (see below).} In other words, a necessary condition for efficient shock acceleration is that $\Delta \theta/\Delta t =  ZeB/\gamma mc \ge \beta^2c/(\beta_S R_S\,\Gamma_S)$. Defining the maximum kinetic energy $E_{max}$ for convenience to be $\sim \beta_{max}^2 \gamma_{max} m c^2$, we may write

\begin{equation}
E_{max}\lesssim Ze\,B\, \beta_S R_S\, \Gamma_S
\label{emax1}
\end{equation}
which generalizes previous results for diffusive shock acceleration  (DSA) [\cite{ei81},  \cite{fordr83} and references therein],  and for scatter-free shock drift, when $\Gamma_S\sim 1$.\footnote{Note that the   expression $E_{max}= Ze\,B\,R_S$, often taken from figure 1 of Hillas (1984), is consistent with equation (\ref{emax1}) only if $\beta_s $ and $\Gamma_s$ are both of order unity.}
Note that scatter-free gyration even in perpendicular shocks, though it gives a much thinner precursor than stochastic propagation,  cannot in general confine a CR particle to a subrelativistic blast wave (e.g. a supernova remnant) in all three dimensions, for the particle would generally drift off to the side within a time $R_s/\beta c$ after gaining the potential difference $Z e B \beta_S  R_S$ in energy.
Also note that we have neglected  adiabatic losses.

{Random scattering, for a given magnetic-field amplitude,
changes the CR's direction more slowly than undisturbed gyration  and, for relativistic shocks, usually  makes it harder for the shock to catch up with a particle. Therefore if
turbulent magnetic field amplification (MFA) increases the field strength to a value $B_{rms}$, equation (\ref{emax1}) should not be used with $B=B_{rms}$   if the coherent scattering angle is less than $1/\Gamma_S$, i.e. if the coherence length of the field $l$ is less than $r_g/\Gamma_S$.


   For random small deflections of $\delta \theta$,  the mean free path $\lambda$ is about $ \beta c/D_{\theta\theta}$, where the angular diffusion coefficient, $D_{\theta\theta}$, is given by $D_{\theta\theta}=(\delta \theta)^2/\delta t \sim [ZeB_{rms}/\beta \gamma mc^2]^2 l^2/(l/\beta c)=r_g^{-2} l \beta c$, where $\delta t \sim l/\beta c$ is the scattering coherence time over which the particle scatters by an angle $\delta \theta$,  $l$ is the coherence length of the magnetic field, and
 $r_g$, in a turbulently enhanced magnetic field, is defined  as  $r_g \equiv \beta \gamma mc^2/ZeB_{rms}$. The  condition for efficient acceleration  is now
  $\overline{\Delta \theta ^2} = D_{\theta \theta}\Delta t
   \ge 1/\Gamma_S^2$,
  or $r_g^2/l \sim \lambda \lesssim \beta_S R_S\Gamma_S^2/\beta$. Finally, we have

\begin{equation}
E_{max}\lesssim Ze\,B_{rms}\,[ l R_S \beta_S\beta]^{\frac{1}{2}}\,\Gamma_S \label{emax2}
\end{equation}
which, for a given field strength, is less than the previous expression for  $ E_{max} $ when $l\le r_g/\Gamma_S$.

This expression for $E_{max}$ implies that, if $l\le r_g/\Gamma_S$, MFA raises $E_{max}$ only if it raises the value of $B_{rms}^2l$. Simply tangling the field on a small scale so that its strength varies as 1/$l^{\eta}$, $\eta \le 1/2$, does not raise $E_{max}$.}

For a self-similar,
energy-conserving relativistic blast wave in the interstellar medium \citep{1976PhFl...19.1130B}
\begin{equation}
 R_S\approx {(17E/16} \pi \rho c^2)^{1/3}\Gamma_S^{-2/3} \approx({ 6  \times 10^{18}}\ {\rm cm})\,
\left(\frac{E_{54}}{n_0}\right)^{1/3}\Gamma_S^{-2/3} .
\end{equation}
{ where the total energy of the blast is $10^{54}E_{54}$ ergs and the ambient nucleon density is $n_0$cm$^{-3}$.}
This suggests, again taking the
 limit for $\Delta \theta/\Delta t$ as arising from coherent
gyration,  that the gyroradius of a maximally energetic escaped particle is
 \begin{equation}
r_{g,max}\approx  \Gamma_S R_S \approx ({ 6 \times 10^{18}}\ {\rm cm})\,
\left(\frac{E_{54}}{n_0}\right)^{1/3}\,\Gamma_S^{1/3}.
\end{equation}

 { In the early stages of a  powerful GRB blast wave, ${ 0.1 \le E_{54}\le 10}$,}   $\Gamma_S\sim 10^3$, while { $10^{-2}\lesssim n_0 \lesssim 1$}. Escaping particles, therefore, should obey  ${ 10^{18.5} \lesssim r_g\lesssim 10^{20.8}}$ cm, and
contribute to the Galactic CR component in the corresponding energy range. They should be represented
in the flux we observe  and { in} the quantity $\dot w_G$, which
is defined below to be the CR power per unit baryon mass within the Galactic
solar circle.

{ The range ${ 10^{18.5}\ {\rm cm}\le r_g \le 10^{20.8}\ {\rm cm}}$
 in the Galaxy corresponds to { the energy range $10^{16}\ {}\,Z\,{ [B/10\mu G]}
\lesssim E\lesssim 10^{18.3}\ {}\,Z\,{ [B/10\mu G]}$} eV,}
precisely the range, to within the
uncertainties, of the "knee-to-ankle" portion of the CR
spectrum, which is said to evade the capabilities of supernovae.
Relativistic blast waves in the Galaxy fill in this range nicely.

Ultrarelativistic shocks are likely  to be quasiperpendicular in the
frame of the shock, and, on these grounds, their ability to accelerate
particles efficiently has been questioned.  We agree  that it is, {\it a
priori}, a fair concern, but note that nonthermal spectra in GRB
afterglows seem to indicate that shock acceleration works there just
fine. The many ways to evade the arguments against shock acceleration
in quasiperpendicular geometries are not the subject of this paper.

\section{The CR power per unit baryon mass}

If we were to suppose that the mechanism supplying
 the sub-ankle CR  somehow extends well beyond the ankle, we would encounter the problem that the Galactic
component of these CR
would be highly anisotropic, assuming their source distribution would be  concentrated inside the solar
circle in the Galaxy, because their transport is no longer fully diffusive and includes
many L\'evy flights. This, in addition to the sharp change in spectral index at the ankle, is  reason to suppose that Galactic GRB limit
their output to CR below the ankle.  Even in the sub-ankle range, the observational
limits on anisotropy pose strong constraints on the models. Below, and in   \citep[Part III: ][]{pe2010}, this is further quantified.

Let $f_{sc,b} M_{sc}=(2\times 10^{44}\ {\rm g})\,f_{sc,b}$ be the total baryonic
mass within the Solar circle, where $M_{sc}\sim 2\times 10^{44}$g.
Using the allsky UHECR integral flux $F{[E_1,E_2]}$  in energy
interval $[E_1,E_2]$ (in units of EeV) implied by Auger, and  assuming that the fluxes
  above the ankle   are
extragalactic and uniform in the cosmos, we find that the UHECR source power
per unit baryon mass in the [4,40] range, $\dot w_{[4,40]}$,   is

\begin{equation}
\dot w_{[4,40]} = \frac{F_{[4,40]}}{\lambda_{[4,40]}\,\Omega _B\,\rho_c}
\approx  (40\ {\rm erg\,g^{-1}\,yr^{-1}})\,
\left({{\lambda_{[4,40]}}\over {\rm Gpc}}\right)^{-1}\, ,
\end{equation}
where $F_{[4,40]}=0.017$ erg/cm$^2$/yr \citep{egp10,abraham}, and  the
implied luminosity from within the Galaxy's solar sphere, $\dot
E_{[4,40]}$, is
\begin{equation}
\dot E_{[4,40]} ={ \dot w_{[4,40],G}}\,M_{sc}\,f_{sc,b}  \approx
(2.5\times 10^{38}\  {\rm erg\,s^{-1}})\,
\left({{\lambda_{[4,40]}}\over {\rm Gpc}}\right)^{-1}\,
f_s^{-1}\,f_{sc,b}, \label{edot}
\end{equation}
Here $\lambda_{[E_1,E_2]}$ is the "horizon" range\footnote{The horizon
range is shorter than the instantaneous range because the expansion of
the universe enhances the losses both by adiabatic deceleration of the
particles and a raising of the background photon energy density and
the losses it causes in the past relative to the present.} of UHECR in
the $[E_1,E_2]$ range, ($\lambda_{[4,40]}\sim 1$~Gpc),
$\Omega _B\rho_c\approx 1.4\times 10^{-31}$g/cm$^3$
is the cosmic density in baryons, and $f_s=
\dot w_{[4,40]}/\dot w_{[4,40,G]} $
is  the ratio of the average UHECR source power
per unit baryon mass to that
in our Galaxy. To be precise, the subscript "G'
denotes the Galactic value within the solar sphere.
Because spiral galaxies like our own comprise about
half the cosmic mass, with the other half in galaxies with less star
formation and hence probably lower UHECR source power, one may estimate the value of
$f_s$ to be about 1/2 if UHECR sources are distributed in proportion to star formation. 


On the other hand, if the solar system fairly samples the outward flux
of  cosmic rays within the energy range ${[E_1,E_2]}$ through the
solar sphere, i.e., if the local flux equals the average over the solar sphere, then the inferred power is given by
\begin{equation}
\dot E_{[E_1,E_2]}=4\pi\,
F_{[E_1,E_2]}\,R_{sc}^2\,\bar \beta_{[E_1,E_2]} \equiv
4\pi\,R_{sc}^2\,\int_{E_1}^{E_2}
E{\bar \beta(E)} cf({ E})d{E}
\end{equation}
where $R_{sc}=8\ {\rm kpc}$ is the radius of the solar circle
and $\bar \beta_{[E_1,E_2]} c$ is the average ratio of  enthalpy flux [in
the anticenter direction, defined to be $\mu\equiv\cos\theta =1$],  $\int_1^2 dE \int d\mu [E \mu f(E,\mu)]$
[where the integral runs from $E_1$ to $E_2$], to
 energy density $4\pi\,\int_1^2 dE [Ef(E)/c]\equiv 2\pi\,\int_1^2 dE \int d\mu [Ef(E,\mu)/c]$. It is measured directly for each energy bin with
CR anisotropy measurements. The current experimental limits on $\bar \beta$ set by the  Auger
Observatory are, to 99\% confidence, $\bar \beta\le$0.004 in the [0.4,4] EeV range and
$\bar \beta \le 0.025$ in the [4,40] EeV range
(Abreu et al., 2011). Here we have used the facts that most of the energy flux is towards the low
end of these ranges, where the limits on anisotropy are strongest, and that   $\bar \beta$ is
1/3  of the first harmonic amplitude given by Abreu et al. (2011).
Under this assumption, we obtain

\begin{equation}
\dot E_{[4,40]}=  F_{[4,40]}\,4\pi\, R_{sc}^2\,\bar\beta_{[4,40]}\lesssim
1\times 10^{35}\ {\rm erg\,s^{-1}}
\label{dotE}
\end{equation}
and correspondingly
\begin{equation}
\dot E_{[0.4,4]}=  F_{[0.4,4]}\,4\pi\, R_{sc}^2\,\bar\beta_{[0.4,4]}
\lesssim 2\times 10^{35}\ {\rm erg\,s^{-1}}
\label{dotE1}
\end{equation}
Equations (\ref{edot}) and (\ref{dotE}), together with the constraints
on $\bar\beta$  imply that
\begin{equation}
f_s=\frac{\dot w_{[4,40]}}{\dot w_{[4,40],G}} \gtrsim 2500 \,f_{sc,b}
\,\left({{\lambda_{[4,40]}}\over {\rm Gpc}}\right)^{-1}
\end{equation}
Note that all UHECR sources that have a power
scaling with star-forming mass, e.g. the hypernova scenario for long GRB, should
have a high likelihood of being present in the Galaxy, i.e. $f_s < 1$.
We conclude that if a) the sources of UHECR are fairly represented in our own Galaxy,
and b) the
solar-system location fairly samples these CR at present, then the
hypothesis that such sources in other galaxies maintain an
extragalactic flux at the observed level would be inconsistent with
the observed CR flux. There would be more CR production
within the solar circle than allowed by observation. This is a
challenge to any theory of UHECR origin from long GRB.

\section{Discussion}

The limit on inferred source power per unit baryon mass required to sustain
Galactic UHECR in the [4-40] EeV range that is imposed by the observed
anisotropy limits is smaller, by more than  3 orders of magnitude,
than what is required for an extragalactic origin, as calculated in \citet{egp10}, and
it corresponds far better to the power per unit mass of gamma rays from GRB.  This numerical
coincidence fits the hypothesis of a GRB origin for  the Galactic component of UHECR,
without invoking  a much larger unseen energy reservoir for GRB. In fact, it would
allow a Galactic origin for UHECR above the ankle were it somehow possible to trap
these CR within the Galaxy effectively enough to obey the isotropy constraint.
It remains to be shown that applying the hypothesis of UHECR from Galactic GRB to
subankle Galactic CR
obeys the
isotropy constraint, and this analysis is done in Part III \citep{pe2010}.

Although the discussion, for historical reasons, has used GRB as a
standard for power production, it is independent  of  GRB. The highest-energy
CRs, whatever their source, are surely extragalactic,
and apparently produced with a higher power per unit (star forming)
mass than that contributed by the matter within the Galactic solar
sphere, given the observed limits of UHECR outflow from this sphere.
This challenges any theory of  their origin from matter and phenomena
of the sort to be found within 10 kpc or so of the Galactic center.

We have considered several alternative possibilities. AGN are an
obvious possibility, as they are not represented by our Galaxy, i.e.
$\dot w/\dot w_G\gg1$.

Another possibility is that the sources are white dwarfs or neutron-star
mergers from binaries in a very extended halo, and that they
spend very little of their time within the solar sphere. Conceivably
this could include short GRB, although their total energy output in
the cosmos is probably an order of magnitude less than even that of long GRB,
so the question of total energetics would still loom large. On the
other hand, short GRB, not being tied to the SFR, need not suffer
the recent decline in rate relative to earlier epochs, and so could be
an order of magnitude more common, relative to long bursts, at present
than in earlier epochs. In any case, one would still have to check
that the implied flux is below observed levels and of a suitable
angular distribution. Why, for example, would there be correlation
above the GZK cutoff with the local supercluster? If one is willing
to attribute sub-GZK CR above the ankle to a different class of
sources from those above the GZK cutoff, then short GRB in the
Galactic halo may account for the former, provided they are distant enough to
respect the strong limits on anisotropy.

In an effort to accommodate the hypothesis of a GRB origin for {\it all} UHECR, we
have also considered the possibility that our present location does
not fairly sample the UHECR exiting the solar sphere, and that the large UHECR output
that would be necessary to supply all of the UHECR at energies where their flux would
be extragalactic  could then mostly evade the solar system.  They conceivably could, for
example, be blown out in jets that have avoided our location and/or
with an intermittency that excluded the present epoch receiving a fair
representation of the time average.
If GRB blasts were to escort all their  CR safely out of the
Galaxy in narrow jets that avoid our location, there would be less of an anisotropy
problem associated with a GRB origin for extragalactic UHECR.
But this scenario would differ from the common view that GRB blasts
slow down to subrelativistic Lorentz factors, spreading in angle,
within the Galaxy.  If the UHECR escape the jet, then according to equation 4, they probably get significantly deflected before escaping the Galaxy at large, and it is not obvious that they could remain sufficiently collimated to conform to an avoidance
scenario.  A single jet, if it leaks UHECR, contaminates the
sky with a strongly anisotropic component at the energy range in which
CR are strongly scattered by the Galactic magnetic field, but not
contained by the jet. If, on the other hand, the jet contains all the
UHECR, then the latter suffer enormous adiabatic losses. The question
is whether all but $\sim10^{-3}$ of the UHECR can avoid
leaking or escaping into the Galaxy at large and mixing with its CR
population. This would appear to require a scenario in which CR at
$\sim10^{19}$eV would be extremely well confined to the shock (strong
scattering) without suffering adiabatic losses and without being
scattered out of the shock's path into the interstellar medium.

Intermittency may explain a low flux from Galactic long GRB,
if the time between GRB per Milky-Way-type galaxy were
more than $R_{sc}/(3c\bar\beta)\simeq 10^4/\bar\beta$ years, the escape time of CR  from the  Galaxy.  The collimation of GRB jets to within several
degrees however,  which is now believed to be the case,  suggests that
GRB are as frequent as $ 10^{-7}f_b^{-1}$ per year per $M_{sc}$, where
$f_b$ is the beaming factor, believed to be of order $10^{-1.5}$  to
$10^{-3}$.  Specifically, the local rate of GRB that we detect is
$R=R_1\ {\rm Gpc^{-3}\,yr^{-1}}$, with $R_1\sim 1$.
The expected  rate $R_{G}$ within the Galaxy should
then be
\begin{equation}
R_G\simeq (1.6\cdot 10^{-9}\ {\rm yr^{-1}})\,\frac{f_{sc,b}\,R_1}
{f_b\,f_s\,\Omega_B}\simeq (2.5\cdot 10^{-5}\ {\rm yr^{-1}})\,
R_1\,f_{sc,b}\,\left(\frac{f_b}{10^{-2.5}}\right)^{-1}\,
\left(\frac{f_s\, \Omega_B}{0.02}\right)^{-1}.
\end{equation}
Given the near isotropy of UHECR, if a good fraction of them were produced within the solar circle, their escape time from the Galaxy, $R_{sc}/\bar\beta c$ would be larger than $1/R_G$, and there would necessarily be many  Galactic GRB per escape time.  Moreover, if the
escape is exponential, more that 10 escape times would be necessary to
clear all but $ 10^{-3}$ of the CR released by the GRB. { These considerations suggest that supply intermittency from GRB is most plausible when  $\bar\beta\sim 1$, but it
is doubtful that the assumption of $\bar\beta\sim 1$ is consistent
with the strong limits on anisotropy.}

On the other hand, the original scenario of \citet{le93}, in which
Galactic GRB supply the Galactic CR component below the ankle, need not
make significant energy demands on the GRB, because there is no
constraint imposed on the required output per unit mass other than
what the local Galactic flux dictates. Given the low  anisotropy of UHECR, it is
likely that those below the ankle are confined to the Galaxy for nearly $10^6$ years, in
which case the rate of GRB and their energy output are consistent with a CR
production per GRB that is less than that of gamma radiation per GRB. This however,
would be incompatible with the hypothesis that GRB produce UHECR above the ankle as well \citep{egp10}.

While the low observed anisotropy eases the energy demands on sources of Galactic UHECR,
it imposes a strong constraint of its own.  It remains to be shown that sources of UHECR,
if distributed as luminous stars in our Galaxy, indeed satisfy the constraint of low anisotropy.
This  question depends on the  question of the mean free paths of UHECR, hence on their composition,  and   is taken up in a companion paper \citep{pe2010}.

Here we have concluded that GRB are energetically sufficient to provide
the sub-ankle Galactic CRs, given
what is known about shock acceleration, relativistic blast waves, and GRB parameters.
If the hypothesis  of GRB origin for sub-ankle UHECR is true, it may have implications for the Galactic magnetic field
and/or the distribution of those GRB.



This research was supported by the Israel-US Binational Science
Foundation, the Israeli Academy of Science, and The Joan and Robert
Arnow Chair of Theoretical Astrophysics.

\end{document}